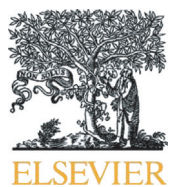

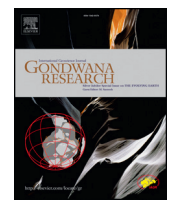

GR Letter

# End-cretaceous cooling and mass extinction driven by a dark cloud encounter

Tokuhiro Nimura [a,*], Toshikazu Ebisuzaki [b], Shigenori Maruyama [c]

[a] Japan Spaceguard Association, 1716-3, Okura, Bisei, Ibara, Okayama 714-1411, Japan
[b] RIKEN, 2-1, Hirosawa, Wako, Saitama 351-0198, Japan
[c] Earth-Life Science Institute, Tokyo Institute of Technology, 2-12-1-IE-1, Ookayama, Meguro-ku, Tokyo 152-8550, Japan

ABSTRACT

We have identified iridium in an ~5 m-thick section of pelagic sediment cored in the deep sea floor at Site 886C, in addition to a distinct spike in iridium at the K–Pg boundary related to the Chicxulub asteroid impact. We distinguish the contribution of the extraterrestrial matter in the sediments from those of the terrestrial matter through a Co–Ir diagram, calling it the "extraterrestrial index" $f_{EX}$. This new index reveals a broad iridium anomaly around the Chicxulub spike. Any mixtures of materials on the surface of the Earth cannot explain the broad iridium component. On the other hand, we find that an encounter of the solar system with a giant molecular cloud can aptly explain the component, especially if the molecular cloud has a size of ~100 pc and the central density of over 2000 protons/cm$^3$. Kataoka et al. (2013, 2014) pointed that an encounter with a dark cloud would drive an environmental catastrophe leading to mass extinction. Solid particles from the hypothesized dark cloud would combine with the global environment of Earth, remaining in the stratosphere for at least several months or years. With a sunshield effect estimated to be as large as −9.3 W m$^{-2}$, the dark cloud would have caused global climate cooling in the last 8 Myr of the Cretaceous period, consistent with the variations of stable isotope ratios in oxygen (Barrera and Huber, 1990; Li and Keller, 1998; Barrera and Savin, 1999; Li and Keller, 1999) and strontium (Barrera and Huber, 1990; Ingram, 1995; Sugarman et al., 1995). The resulting growth of the continental ice sheet also resulted in a regression of the sea level. The global cooling, which appears to be associated with a decrease in the diversity of fossils, eventually led to the mass extinction at the K–Pg boundary.

© 2016 The Authors. Published by Elsevier B.V. on behalf of International Association for Gondwana Research. This is an open access article under the CC BY license (http://creativecommons.org/licenses/by/4.0/).

## 1. Introduction

The Milky Way Galaxy has a disk with a radius of 10 kpc and a thickness of 200 pc. Our solar system is located inside at ~8.5 kpc from its center. Many dark clouds are distributed in the disk.

A dark cloud consists of high-density (100–10,000 protons/cm$^3$) (Goldsmith, 1987) and low-temperature (8–40 K) (Goldsmith, 1987) neutral gas. The size of dark clouds ranges ~0.2–174 pc (Dame et al., 1986; Goldsmith, 1987; Solomon et al., 1987). It has been estimated that ~135 clouds of over ~100 protons/cm$^3$ and ~16 clouds of over ~1000 protons/cm$^3$ encounter likely occurs in the age of the sun ($4.6 \times 10^9$ years) (Talbot and Newman, 1977). Maruyama and Santosh (2008) suggested that some Earth's dramatic environmental changes are controlled by trigger in outer the Earth, solar system, or/and galaxy. A number of previous studies have suggested that a nebula encounter may lead to an environmental catastrophe (Whitten et al., 1963; Ruderman, 1974; Begelman and Rees, 1976; Clark et al., 1977; Talbot and Newman, 1977; Pavlov et al., 2005a, 2005b). Kataoka et al. (2013,

2014) pointed that the encounter with a dark cloud may drive environmental catastrophe to lead a mass extinction. A nebula encounter leads to a global cooling by the enhanced flux of cosmic dust particles. They remain in the stratosphere for several months or even years and work as sunscreens to reflect back the sunlight.

The evidence of a dark nebula encounter can be found in deep sea sediments, as an iridium-rich layer, if the sedimentation speed is less than several m Myr$^{-1}$. The iridium-rich dust flux from a dark cloud on the surface of the Earth is estimated to be as high as 1 kg m$^{-2}$ Myr$^{-1}$ in the case of a gas density of ~10$^3$ protons/cm$^3$ and a relative velocity of 10 km s$^{-1}$. The iridium abundance in the sediment is inversely proportional to the sedimentation speed, $v$ (Fig. 1), and can be as high as 0.05 ppb in the case of deep-sea sediment, where the sedimentation speed is as low as 3 m Myr$^{-1}$. This level of iridium density is detectable through radiochemical neutron activation analysis, with the element's detection limit being typically 0.02–0.05 ppb (Kyte et al., 1995). The accumulation of iridium in the sediment is estimated to take ~10 Myr, with a relative velocity of 10 km s$^{-1}$ typical relative velocity for dark clouds relative to Earth at galactic longitude is 270° (Dame et al., 2001; Heyer and Dame, 2015), since the sizes of giant molecular clouds are as large as 100 pc.

Through the Ocean Drilling Program (ODP), a core sample of pelagic sediment was taken at Site 886C, at a water depth of 5713.3 m in the central North Pacific (44°41.384′N, 168°14.400′W) (Fig. 2). This site has been in an area of pelagic accumulation at least since the end of the Cretaceous Period. The accumulation rate is estimated to be as low as 3 m Myr$^{-1}$ (Fig. 1) (Ingram, 1995). Kyte et al. (1995) measured the iridium density in the core sample of Site 886C at a depth of 0.75–72.2 m below the sea floor (mbsf) with a pelagic-sedimentation record extending from the Late Cretaceous (77.77 Ma at 71.60 mbsf) to the Late Miocene (9.8 Ma at 54.6 mbsf). The age of sediment was determined based on Sr-isotope stratigraphy below 54 mbsf of 886C (Ingram, 1995).

* Corresponding author.
E-mail addresses: nimura@spacegurd.or.jp, tokuhiro.nimura@riken.jp (T. Nimura).

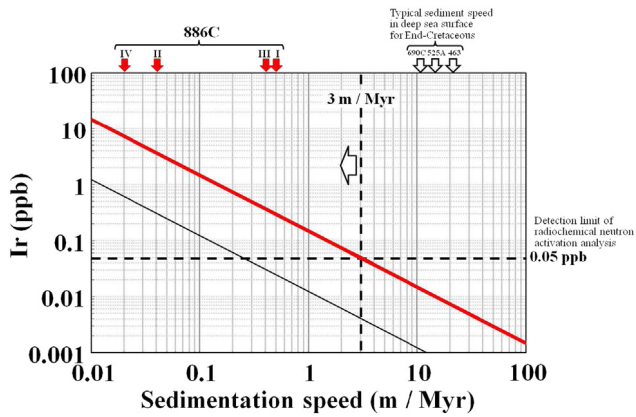

**Fig. 1.** Relationship between iridium abundance and sedimentation speed. Whereas the thin black solid line delineates the present interplanetary dust particles (IDP; $4 \times 10^4$ tons yr.$^{-1}$ (Pavlov et al., 2005b)), as a function of sedimentation speed, the thick red solid line defines an enhanced IDP; corresponds to the dark cloud encounter with density is $10^3$ protons/cm³, and relative velocity of solar system and cloud is 10 km s$^{-1}$, while the horizontal dash line defines the iridium detection limit from radiochemical neutron activation analysis (0.02–0.05 ppb) (Kyte et al., 1995). The iridium anomaly is detectable, only where the sedimentation speed is lower than ~3 m Myr$^{-1}$ (vertical dash line). We found that the four periods of the Site 886C (I, II, III, and VI) are above the detection limit (horizontal dash line). On the other hand, the average sedimentation speeds of Site 463 (Thiede et al., 1981), 525 A (Shackleton et al., 1990), and 690C (Stott and Kennett, 1990) in End-Cretaceous turns out of less than the detection limit, therefore the detection of encounter components is difficult there, because of the dilution by the Earth's materials to be detected. (For interpretation of the references to color in this figure legend, the reader is referred to the web version of this article.)

## 2. Co–Ir diagram

The iridium abundance is plotted against cobalt abundance in Fig. 3, normalized by those of CI chondrite. The large open squares represent Earth's surface materials (Table 1) (i.e., marine organisms (Martic et al., 1980; Wells et al., 1988), earth crust (Taylor and McLennan, 1985), pelagic clays (Goldberg et al., 1986; Terashima et al., 2002), and manganese nodule (Harriss et al., 1968; Glasby et al., 1978; Usui and Moritani, 1992)). The large solid squares represent CI chondrite (Lodders, 2003). The cobalt abundance of pelagic clays is used 222 ± 41 ppm (Terashima et al., 2002) as similar to 886C sample which are slow sedimentation rate (<1 m/Myr) and deep (>5000 mbsf) sea sediment. The positive and negative error value for iridium and cobalt abundance for crust are adopted upper and lower crust (Taylor and McLennan, 1985). Any mixtures of the Earth's surface materials must distribute in the gray area taken into account the measurement errors ($3\sigma$ in iridium abundance. The standard deviation of the iridium is estimated by the data in 0.75–54.4 mbsf of the core.) and statistical error. The statistical error is estimated by Table 1. These values of the uncertainty are consistent with the detection limited reported in Kyte et al. (1995). We found that most of the data points in 886C are located well above the curves.

Here, the iridium abundance of a mixture of Earth's surface material is calculated through three linear combinations of materials of Earth, including marine organisms (MO), marine crust (MC), pelagic clays (PC), and manganese nodule (MN), with the contribution of exosolar material ($Y_{EX}$) as

$$
\begin{aligned}
Y_{EX} &= Y - \left\{ \left(3.0 \times 10^{-6}\right)X + \left(1.4 \times 10^{-5}\right) \right\} \text{ for MO to MC,} \\
&= Y - \left\{ \left(2.1 \times 10^{-6}\right)X + \left(4.0 \times 10^{-5}\right) \right\} \text{ for MC to PC,} \\
&= Y - \left\{ \left(2.6 \times 10^{-6}\right)X - \left(7.3 \times 10^{-5}\right) \right\} \text{ for PC to MN,}
\end{aligned} \tag{1}
$$

where $Y$ and $X$ are iridium and cobalt abundances in ppm. Assuming that the excess iridium abundance in the sediment originates from exosolar materials, the extraterrestrial index $f_{EX}$ can be defined as

$$
f_{EX} = v \; q \; \left( \frac{Y_{EX}}{Y_{CI}} \right) \tag{2}
$$

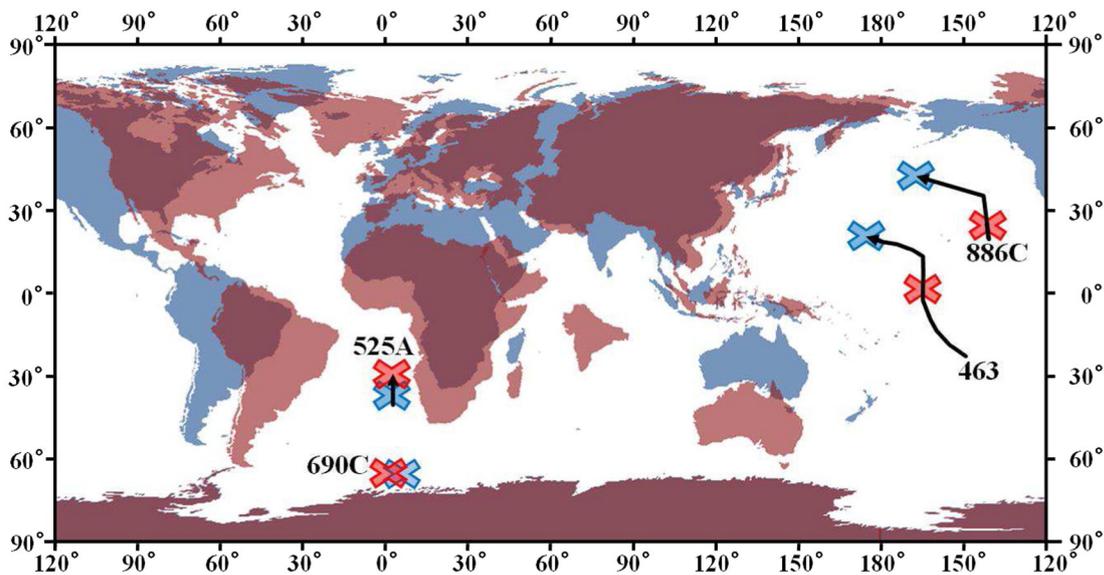

**Fig. 2.** Location of Sites 463, 525 A, 690C, and 886C. Configuration of continents and site locations at 65 Ma (red) and at present-day (blue), with solid lines delineating travel paths (Thiede et al., 1981; Snoeckx et al., 1995; Li and Keller, 1999). An iridium anomaly is caused by extraterrestrial material and concentration of iridium-rich material of Earth (e.g., manganese nodule). These extraterrestrial and terrestrial factors can be distinguished by multi-component analysis through a Co–Ir diagram as discussed below. (For interpretation of the references to color in this figure legend, the reader is referred to the web version of this article.)

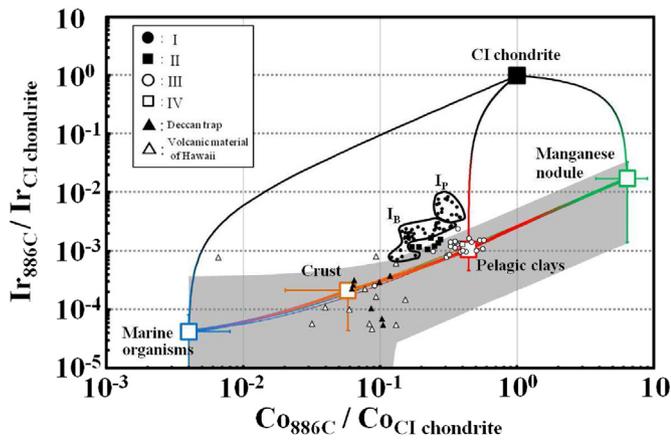

**Fig. 3.** A Co–Ir diagram detailing iridium abundances plotted against cobalt abundance, with both normalized to CI-chondrite. The different symbols correspond to the data of periods listed in Table 2. The large open squares represent marine organisms (blue), earth crust (brown), pelagic clays (red), and manganese nodule (green), respectively. The large black solid square represents CI chondrite. The mixtures of Earth's surface material must be distributed in the gray area. The filled and open triangles represent the data of the Deccan Traps (Shukla et al., 2001) and volcanic material of Hawaii (Crocket, 2002), respectively. (For interpretation of the references to color in this figure legend, the reader is referred to the web version of this article.)

where $v$, $q$, and $Y_{CI}$ are sedimentation speed, a density of the core sample, and the iridium abundance of CI chondrite, respectively. The index $f_{EX}$ gives the flux of extraterrestrial matter at the time of the sedimentation.

## 3. Extraterrestrial index ($f_{EX}$)

Fig. 4b shows the iridium abundance of the core sample of Site 886C (Kyte et al., 1995). The core sample can be divided into four periods (I–IV) in terms of the sedimentation rates (see corresponding Fig. 4e and Table 2). Period I (solid circles in Fig. 3 and Table 2, 64.10–72.20 mbsf) with a sedimentation rate of 0.5 m Myr$^{-1}$ spans the Campanian, Danian, and Masstrichtian ages (61.6–72.1 Ma). There is a sharp spike at 65.24 mbsf (solid arrow; component $I_P$) with a maximum value of 4.01 ppb which corresponds to the Chicxulub asteroid impact (Kyte et al., 1995). There is also a broad component (component $I_B$) which has iridium abundance as high as 0.3 ppb. Since the solid circles (period I) locate significantly above the gray area in the Co–Ir diagram (Fig. 3), their iridium abundance cannot be explained by any mixture of the materials on the surface of the Earth, and thus requires contribution from extraterrestrial materials, such as period $I_P$ and $I_B$. Note that the possibility of the contribution of manganese nodules can be excluded, if we take into account that manganese nodules are abundant both in cobalt and iridium. Even for the component $I_B$, the extraterrestrial flux is required to be larger than $1.3 \times 10^6$ tons yr$^{-1}$, which is 33 times larger than the present interplanetary dust particle (IDP) flux $4 \times 10^4$ tons yr$^{-1}$ (Pavlov et al., 2005b). This enhanced flux of the extraterrestrial material can be explained by a dark cloud encounter as discussed later.

It is worth noting that this iridium-enriched zone continues over 5 m, which corresponds to the ages from Campanian to Danian. Neither diffusion nor bioturbation can explain this broad anomaly because of the following reasons: (1) redistribution of platinum-group element such as Pt, Re, and Ir is generally less than 10 cm (Colodner et al., 1992); (2) the mean depth of marine bioturbation is as small as ~10 cm (Boudreau, 1998), which is much less than the width of the broad component (~5 m), and (3) the evidence of bioturbation is minor in the core.

Period II (small solid squares in Fig. 3 and Table 2, 62.85–64.10 mbsf) covers Danian age (61.6–66.0 Ma) to Ruperian age (28.1–33.9 Ma). The sedimentation rate is 0.04 m Myr$^{-1}$, which is the second slowest among the four periods. The features in this period plot significantly above the gray area. This excess in iridium can be explained by the present IDP flux. In other words, no extra IDP flux is necessary.

The sedimentation rate of Period III (open circles in Fig. 3 and Table 2, 58.60–62.85 mbsf) is estimated to be 0.4 m Myr$^{-1}$, covering Ruperian age (28.1–33.9 Ma) and Aquitanian age (20.44–23.03 Ma). In the Co–Ir diagram (Fig. 3), the open circles are plotted in the gray area. The iridium abundance of this period can be explained by the mixtures of the materials on the surface of the Earth.

The period IV (small open squares in Fig. 3 and Table 2, 58.35–58.06 mbsf) covers Aquitanian age (20.44–23.03 Ma) and Tortonian age (7.246–11.63 Ma). The sedimentation rate is 0.02 m Myr$^{-1}$, which is the slowest among the four periods.

In summary, Period I requires an enhanced flux of extraterrestrial materials probably caused by an encounter with a dark cloud to explain the relatively high iridium abundance, while other periods can be explained by a mixture of surface materials of Earth, if the present level of IDP is taken into account.

Officer and Drake (1985) argued that the iridium anomaly can be explained by a mixture of volcanic basalt like Kilauea. However, the Co–Ir diagram (Fig. 3) shows that the data points of the broad component of period I ($I_B$, solid circle) are distributed in a region different from those of the Decan Trap (solid triangle of Fig. 3) and Hawaii (open triangle of Fig. 3). The volcanic basalts, which locate in the gray area cannot produce materials outside the area.

## 4. Dark cloud encounter

Fig. 5a shows the estimated flux of exosolar material of 10–80 Ma through (1) and (2). We assumed a density of the core sample to be 3 g cm$^{-3}$. The enhanced flux of exosolar material began ~73 Ma and has continued to fall through ~8 Myr. The peak at 66 Ma is likely to be originated from an asteroid impact which formed the Chicxulub impact crater (Kyte et al., 1995). On the other hand, the broad component can be explained not solely by the Chicxulub impact but by an encounter with a giant molecular cloud. The encounter leads to a "Nebula Winter" (Kataoka et al., 2013, 2014), in which an environmental catastrophe of the Earth is driven by an enhanced flux of cosmic dust particles and cosmic rays, which cause global cooling and destruction of the ozone layer (Fig. 6).

The scale on the right hand side of Fig. 5a denotes the density of nebula $N$, corresponding to dust flux in the case of relative velocity of the solar system and cloud $V$ is 10 km s$^{-1}$ and 20 km$^{-1}$ by following equation:

$$N = 1000 \text{ protons } / \text{ cm}^3 \left[ \frac{f_{EX}}{1.0 \times 10^{-8} \text{kg m}^{-2} \text{ yr}^{-1}} \right] \left[ 1.0 \left( \frac{V}{20 \text{ km s}^{-1}} \right) \right. \quad (3)$$
$$\left. + 4.6 \left( \frac{V}{20 \text{ km s}^{-1}} \right)^{-1} \right]^{-2}.$$

The dark cloud density of 2000–6000 protons/cm$^3$ corresponds to the climate forcing ($F$) of snowball earth ($-9.3$ W m$^{-2}$ (Pavlov et al., 2005b)), depending on the relative velocity (Fig. 5), for the case that typical size ($a$) and density ($\rho$) of cosmic dust particles are 0.2 μm and 3 g cm$^{-3}$ by following equation (Kataoka et al., 2013):

$$f_{EX} = \left(3.1 \times 10^{-7} \, kg \, m^{-2} \, yr^{-1}\right) \left(\frac{F}{-20 \, W \, m^{-2}}\right) \left(\frac{a}{0.1 \, \mu m}\right)^{2.3} \left(\frac{\rho}{10^3 \, kg \, m^{-3}}\right). \tag{4}$$

| | Ir (ppb) | Co (ppm) |
| --- | --- | --- |
| Marine organism | $0.02 \pm 0.02$ | $2 \pm 2$ |
| Crust | 0.1 (0.02–0.13) | 29 (10–35) |
| Pelagic clay | $0.5 \pm 0.3$ | $222 \pm 41$ |
| Manganese nodule | $8.18 \pm 7.50$ | $3200 \pm 1300$ |
| CI chondrite | $470 \pm 5$ | $502 \pm 17$ |

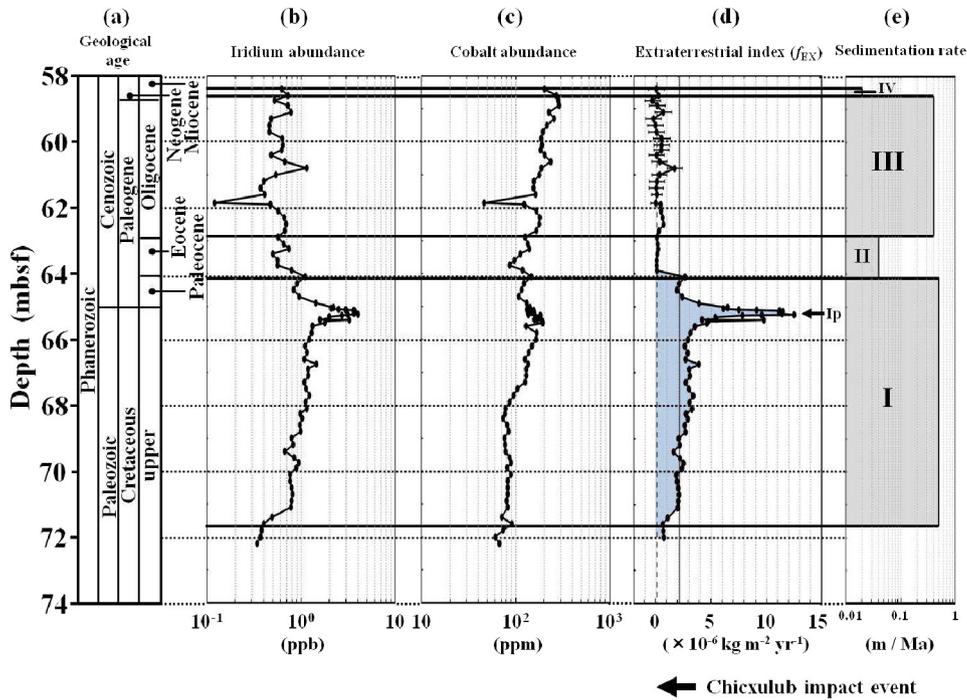

**Fig. 4.** Geological age (a), iridium abundance (b) (Kyte et al., 1995), cobalt abundance (c) (Kyte et al., 1995), extraterrestrial index (d), and sedimentation rate (e) (Snoeckx et al., 1995), as well as the pelagic deep sea sediments sampled at 886C, marking the End-Cretaceous around the K–Pg boundary. The blue area cannot be explained by any mixture of the surface materials of the Earth. (For interpretation of the references to color in this figure legend, the reader is referred to the web version of this article.)

The nebula density, which is the density about $10^3$ protons/cm$^3$, can be seen in a typical dark nebula core. It is worth noting that the optical depth is given as (Kataoka et al., 2013).

$$\tau = 1.1 \times 10^{-2} \left(\frac{f_{EX}}{3.1 \times 10^{-7} \, kg \, m^{-2} \, yr^{-1}}\right) \left(\frac{t}{10 \, yr}\right) \left(\frac{a}{0.2 \, \mu m}\right)^{-1} \tag{5}$$
$$\times \left(\frac{\rho}{10^3 \, kg \, m^{-3}}\right)^{-1},$$

**Table 2**
Summary of the analysis of the core sample of 886C and volcanic materials of Decan Trap and Hawaii.

| Symbol | Period | | Depth (mdsf) | | Age (Ma) | | Sedimentation speed (m/Myr) | Putative origin of iridium | Extraterrestrial flux (×present IDP) |
|---|---|---|---|---|---|---|---|---|---|
| | | | min | max | min | max | | | |
| □ | IV | | 58.35 | 58.60 | 11.61 | 22.27 | −0.02 | Terrestrial | 0.31 |
| ○ | III | | 58.60 | 62.85 | 22.27 | 33.64 | −0.4 | Terrestrial | – |
| ■ | II | | 62.85 | 64.10 | 33.64 | 63.87 | −0.04 | Usual IDP | 1.4 ± 0.7 |
| ● | I | B | 64.10 | 65.03 | 63.87 | 65.59 | −0.5 | Extraterrestrial | 33 ± 10 |
| | | P | 65.03 | 65.50 | 65.59 | 66.46 | | | 54 – 159 |
| | | | 65.50 | 71.60 | 66.46 | 77.77 | | | 33 ± 8 |
| | | B | 71.60 | 72.20 | 77.77 | – | – | | 8.3 ± 0.7 |
| ▲ | | | | | | | | Deccan trap | |
| △ | | | | | | | | Volcanic material of Hawaii | |

where $t$ is the residence time in the stratosphere estimated as (Kataoka et al., 2013).

$$t = 18 \, yr \left(\frac{a}{0.2 \, \mu m}\right)^{-1.3}. \tag{6}$$

The optical depth is as high as ~0.045 (Fig. 5a).

Fig. 6 depicts the concept of the encounter with the dark cloud. During the encounter of a dark cloud, the fluxes of cosmic dust particles and cosmic rays are enhanced. Both of them lead to global climate cooling. The following three evidences show that global cooling took place well before K–Pg boundary (~73–65 Ma).

First, the oxygen isotope ratio shows strong climate cooling. The lighter oxygen $^{16}$O tends to be contained more in water vapor than in heavier $^{18}$O. This lighter water vapor will settle down to the land as snow to form massive ice sheets on the continents. Therefore, the lower $\delta^{18}$O is incorporated in the ice sheets on the continents, while the higher $\delta^{18}$O remains in the ocean. The shell of planktonic and benthic foraminifera is a good indicator of oceanic $\delta^{18}$O. Their dead bodies settle along the ocean floor to form a sediment layer. Therefore, the positive anomaly in oxygen isotope ratio (Fig. 7) (Barrera and Huber, 1990; Li and Keller, 1998; Barrera and Savin, 1999; Li and Keller, 1999) can be considered as a result of the extension of the continental ice sheet, which suggests global cooling during the End-Cretaceous from ~73 Ma. The trends of $\delta^{18}$O can also be consistently seen in low-, mid-, and high-latitude ODP samples (Site 463, 525 A, and 690C).

Second, global cooling is also suggested by high $^{87}$Sr/$^{86}$Sr ratios. The development of an ice sheet leads to both marine regression and exposure of continental shelves. The resultant enhancement of denudation and weathering leads to a high $^{87}$Sr/$^{86}$Sr because of a high $^{87}$Sr/$^{86}$Sr in the continental crust. The increase in $^{87}$Sr/$^{86}$Sr during the End-Cretaceous (Fig. 8) (Barrera and Huber, 1990; Ingram, 1995; Sugarman et al., 1995) suggests an increase in denudation and weathering beginning at ~73 Ma and continuing for ~8 Myr due to the lower sea level, which indicates a globally cool climate. The trends of $^{87}$Sr/$^{86}$Sr can also be consistently seen in low-, mid-, and high-latitude ODP samples (Site 463, 525 A, and 690C). Finally, the sea-level changes in End-Cretaceous (Hallam and Wignall, 1999) can be caused by an extension of the continental ice sheet.

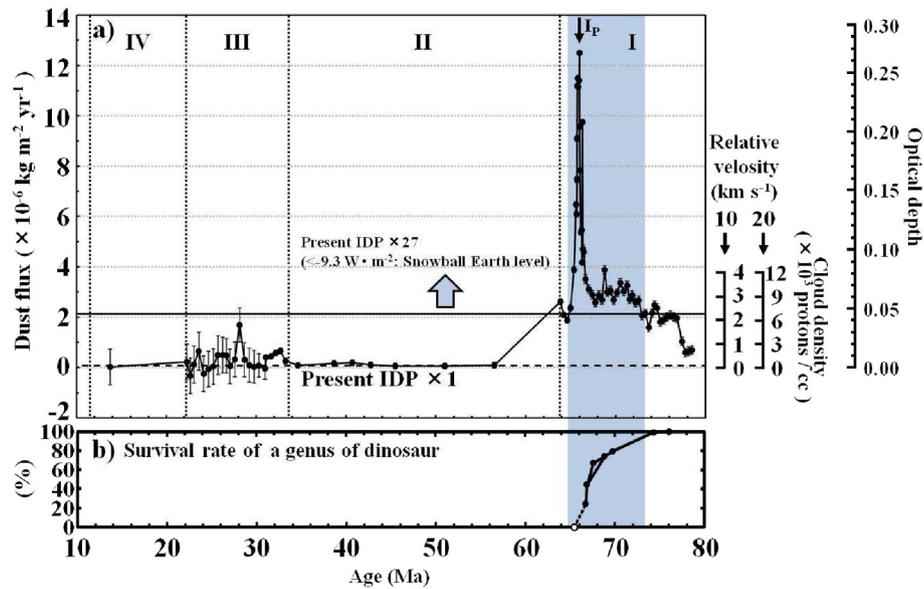

**Fig. 5.** a) Dust flux and dark cloud density, which are estimated through Eqs. (1)–(3), are shown. Whereas the horizontal axis represents geological age (Snoeckx et al., 1995), the vertical axis is the density of the dark cloud in the case of the velocity of the solar system relative to the dark cloud is 10 km s$^{-1}$ and 20 km s$^{-1}$, and optical depth of the dust particles in the stratosphere. b) The survival rate of a genus of dinosaur (Sloan et al., 1986). The blue area delineates 73–65 Ma, a period of time when there is recorded enhanced climate cooling hypothesized to be the result of the dark cloud encountering the terrestrial environment, with the dinosaurs gradually becoming extinct. (For interpretation of the references to color in this figure legend, the reader is referred to the web version of this article.)

The long period of climate cooling forced by a dark-cloud encounter may account for a mass extinction at the end of the Cretaceous. It is also consistent with a trend of climate cooling in generic reduction of dinosaur (Sloan et al., 1986) (Fig. 5b), as well as ammonite (Hancock, 1967; Wiedmann, 1973; House, 1989, 1993).

## 5. Chicxulub impact at K–Pg boundary

The mass extinction at the K–Pg boundary is widely thought to have been caused by an impact of an asteroid (e.g., Alvarez et al., 1980; Schulte et al., 2010) at 65.5 Ma. However, a complete extinction of the total family by just one asteroid impact seems rather difficult because of the following two reasons. First, a severe perturbation in the global environment of Earth would finish nearly 5 years after the impact

(Planetary Science Institute, 2015), the solid particles (including sulphate) launched by the asteroid impact would settle down in only a few months (troposphere) to a few years (stratosphere), and a negative radiative forcing would become negligible a few years subsequent to the impact. For example, the effect of the sunshield by launched solid particles and sulphate into the atmosphere of the Earth from the eruptions of the Pinatubo (June 1991, 20 Mt) and Toba (about 71 kyr ago, 1 Gt) volcanoes completely vanished after 3–5 years (Planetary Science Institute, 2015). Even if 100 times larger than the Toba eruption, as is the estimated case of the Chicxulub impact, the sunshield effect would not persist beyond about 5 years, because the retention time is determined by convection time of stratosphere. Individuals of those few percentage of surviving species would recover completely after the environmental catastrophe was over.

Second, there have been impact events of similar scale to that of Chicxulub without catastrophe of the global environment/biosphere, such as the Woodleigh, Chesapeake Bay, and Popigai impact cratering ($\sim 2 \times 10^{15}$–$4 \times 10^{16}$ g) is similar to that determined for the Chicxulub crater ($10^{14}$–$10^{16}$ g) (Kring et al., 1996), though no mass extinction took place. Therefore, why only the Chicxulub impact would lead to mass extinction but not the other three comparable impacts is difficult to explain by an impact hypothesis alone.

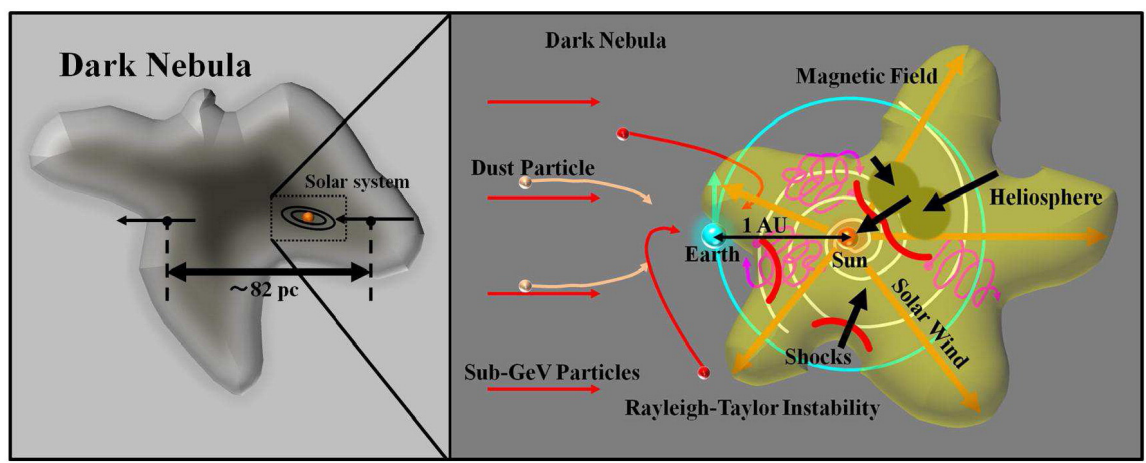

**Fig. 6.** Concept of the encounter of the terrestrial environment with a dark cloud of the size of ~80 pc. The heliosphere shrinks down to ~1 AU from the current size (~400 AU). The cloudlets formed by Rayleigh–Taylor instability collide with each other in the inter-planetary space to form ~GeV cosmic rays (Kataoka et al., 2013).

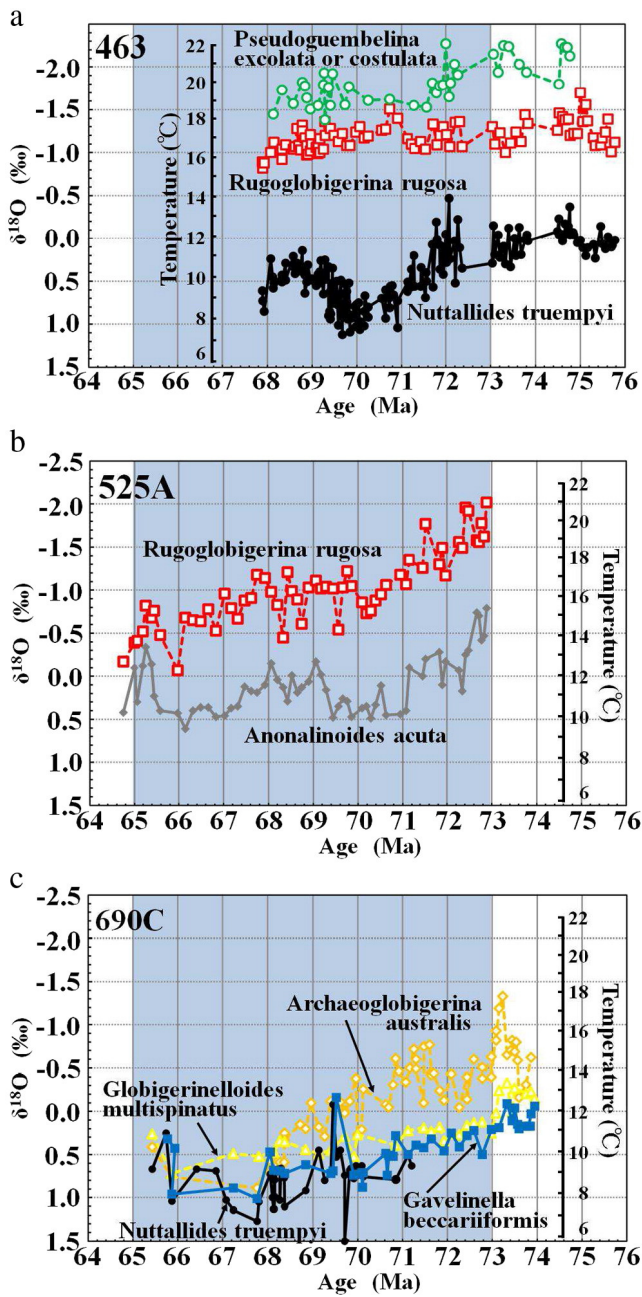

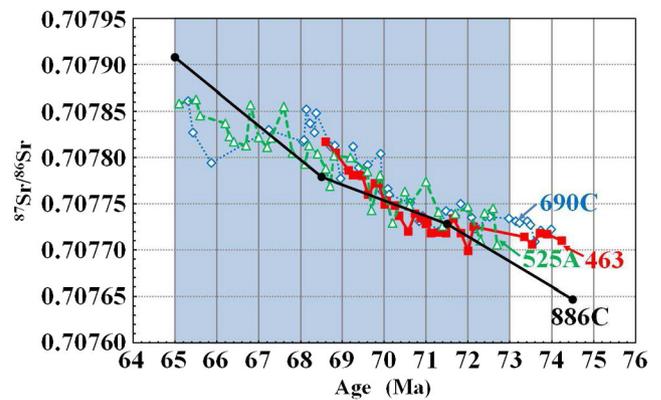

**Fig. 8.** Strontium isotope ratio ($^{87}$Sr/$^{86}$Sr) based on low-, mid-, and high-latitude ODP samples (463, 525 A, 690C, and 886C) (Barrera and Huber, 1990; Ingram, 1995; Sugarman et al., 1995) for the latest Cretaceous.

It is worth noting that an encounter with the dark cloud perturbs the orbit of asteroids and comets through its gravitational potential, causing showers of asteroids and/or comets, and likely associated impact events (Kataoka et al., 2014). The asteroid impact at K–Pg, therefore, may be one of the consequences of the dark cloud encounter. On the other hand, the broad component that span more than 8 Myrs is unlikely to be caused by a collision in the asteroid belt, since the fall down time scale of 0.2 μm dust from the asteroid belt ($\sim 3$ AU) is as short as $\sim 6000$ yrs. because of Poynting Robertson effect (Burns et al., 1979).

We conclude that the cause of the cold climate at the End-Cretaceous was driven by an encounter with a giant molecular cloud, with such an encounter and related perturbation in global climate a more plausible explanation for the mass extinction than a single impact event, Chicxulub.

## 6. Conclusion

We found that the multi-component analysis and "extraterrestrial index" $f_{EX}$ are effective to show the evidence of the abnormal extraterrestrial flux in the past Earth from deep-sea sediments. This method can be applied (and further tested) to other periods of drastic change in the Earth's environment such as End-Triassic, End-Permian, F–F boundary, and End-Ordovician events, recorded in the sedimentary records of accretionary prisms. The oxygen and strontium isotope ratios are consistent with the gradual cooling of the End-Cretaceous period due to the encounter with a giant molecular cloud. It is also consistent with the gradual decrease of the number of genera. If this hypothesis is correct, then other index elemental information (e.g., plantinum group elements and stable isotope ratios) and the distribution of spherules of exosolar origin will coeval with the broad component of the iridium anomaly ($I_B$). The spherules will be found as investigation proceeds of deep sea sediments. A glacial deposit such as tillitise and/or dropstone are expected to be found.

## Acknowledgments

This work was supported by JSPS KAKENHI grant numbers 26106002, 26106006. T.N. wants to thank Yumi Awano who is director of Okayama Astronomical Museum. James Dohm gives constructive comments and suggestions.

**Fig. 7.** Oxygen isotope ratio based on low-, mid-, and high-latitude ODP samples (Site 463, 525 A, and 690C) (Barrera and Huber, 1990; Li and Keller, 1998; Barrera and Savin, 1999; Li and Keller, 1999) for the latest Cretaceous. The dashed and solid lines represent planktonic and benthic foraminifera, respectively. A model temperature was calculated using T = 16.28 − 4.47($\delta_c$ + 1) + 0.109($\delta_c$ + 1) (Barrera and Savin, 1999), where $\delta_c$ is the $\delta^{18}$O value of the foraminifer CaCO$_3$ relative to Pee Dee Belemnite. However, the temperatures, which are calculated using the $\delta^{18}$O value of benthic foraminiferal species, are about one degree lower than this model temperature.

events. In particular, there is no discernable evidence for a Late Devonian (364 Ma) mass extinction related to the formation of the 120-km-diameter Woodleigh crater in western Australia (Mory et al., 2000a, 2000b; Uysal et al., 2001; Glikson et al., 2005), about ~80% the size of the Chicxulub crater (~150 km diameter) (Earth Impact Database, 2015). The 85-km-diameter Chesapeake Bay (Poag et al., 2002) and the 100-km-diameter Popigai (Earth Impact Database, 2015) impact craters were formed during Late Eocene epoch in America and Russia, respectively. Also respectively, the diameters of these craters are ~57% and ~67% of Chicxulub crater. Furthermore, in the case of the Popigai crater, the stratospheric injection of impact-produced sulfur

## References

Alvarez, L.W., Alvarez, W., Asaro, F., Michel, H.V., 1980. Extraterrestrial cause for the Cretaceous-Tertiary extinction. Science 208, 1095–1108.

Barrera, E., Huber, B.T., 1990. Evolution of antarctic waters during the maestrichtian: foraminifer oxygen and carbon isotope ratios, leg 113. Proceeding of the Ocean Drilling Program, Scientific Results 113, 813–827.

Barrera, E., Savin, S.M., 1999. Evolution of late Campanian–Maastrichtian marine climates and oceans. Geological Society of America Special Paper 332, 245–282.

Begelman, M.C., Rees, M.J., 1976. Can cosmic clouds cause climatic catastrophes? Nature 261, 298–299.

Boudreau, B.P., 1998. Mean mixed depth of sediments: the wherefore and the why. Limnology and Oceanography 43, 524–526.

Burns, J.A., Lamy, P.L., Soter, S., 1979. Radiation forces on small particles in solar system. Icarus 40, 1–48.

Clark, D.H., McCrea, W.H., Stephenson, F.R., 1977. Frequency of nearby supernovae and climatic and biological catastrophes. Nature 265, 318–319.

Colodner, D.C., Boyle, E.A., Edmond, J.M., Thomson, J., 1992. Post-depositional mobility of platinum, iridium and rhenium in marine sediments. Nature 358, 402–404.

Crocket, J.H., 2002. Platinum-group elements in basalts from Maui, Hawaii: low abundances in alkali basalts. The Canadian Mineralogist 40, 595–609.

Dame, T.M., Elmegreen, B.G., Cohen, R.S., Thaddeus, P., 1986. The largest molecular cloud complexes in the first galactic quadrant. The Astrophysical Journal 305, 892–908.

Dame, T.M., Hartmann, D., Thaddeus, P., 2001. The milky way in molecular clouds: a new complete CO survey. The Astrophysical Journal 547, 792–813.

Earth Impact Database, 2015. http://www.passc.net/EarthImpactDatabase/.

Glasby, G.P., Keays, R.R., Rankin, P.C., 1978. The distribution of rare earth, precious metal and other trace elements in recent and fossil deep-sea manganese nodules. Geochemical Journal 12, 229–243.

Glikson, A.Y., Mory, A.J., Iasky, R.P., Pirajno, F., Golding, S.D., Uysal, I.T., 2005. Woodleigh, southern Carnarvon basin, western Australia: history of discovery, late devonian age, and geophysical and morphometric evidence for a 120 km-diameter impact structure. Australian Journal of Earth Sciences 52, 545–553.

Goldberg, E.D., Hodge, V.F., Kay, P., Stallard, M., Kolde, M., 1986. Some comparative marine chemistries of platinum and iridium. Applied Geochemistry 1, 227–232.

Goldsmith, P.F., 1987. Molecular clouds: an overview. In: Hollenbach, D.J., Thronson Jr., H.A. (Eds.), Interstellar Processes. Proceedings of the Symposium on Interstellar Processes, Held in Grand Teton National Park, July 1986. Astrophysics and Space Science Library 137, pp. 51–70.

Hallam, A., Wignall, P.B., 1999. Mass extinctions and sea-level changes. Earth-Science Reviews 48, 217–250.

Hancock, J.M., 1967. Some Cretaceous–Tertiary marine faunal changes. Geological Society of London, Special Publications 2, 91–104.

Harriss, R.C., Crocket, J.H., Stainton, M., 1968. Palladium, iridium and gold in deep-sea manganese nodules. Geochimica et Cosmochimica Acta 32, 1049–1056.

Heyer, M., Dame, T.M., 2015. Molecular clouds in the milky way. Annual Review of Astronomy and Astrophysics 53, 585–629.

House, M.R., 1989. Ammonoid extinction events. Philosophical Transactions of the Royal Society of London B325, 307–326.

House, M.R., 1993. Fluctuations in ammonoid evolution and possible environmental controls. In: House, M.R. (Ed.), The Ammonoidea: environment, ecology, and evolutionary change The Systematics Association Special 47. Clarendon Press, Oxford, pp. 13–34.

Ingram, B.L., 1995. Ichthyolith strontium isotopic stratigraphy of deep-sea clays: sites 885 and 886 (North Pacific transect). Proceeding of the Ocean Drilling Program, Scientific Results 145, 399–412.

Kataoka, R., Ebisuzaki, T., Miyahara, H., Maruyama, S., 2013. Snowball earth events driven by starbursts of the milky way galaxy. New Astronomy 21, 50–62.

Kataoka, R., Ebisuzaki, T., Miyahara, H., Nimura, T., Tomida, T., Sato, T., Maruyama, S., 2014. The Nebula Winter: the united view of the snowball earth, mass extinctions, and explosive evolution in the late Neoproterozoic and Cambrian periods. Gondwana Research 25, 1153–1163.

Kring, D.A., Melosh, H.J., Hunten, D.M., 1996. Impact-induced perturbations of atmospheric sulfur. Earth and Planetary Science Letters 140, 201–212.

Kyte, F.T., Bostwick, J.A., Zhou, L., 1995. Identification and characterization of the Cretaceous/Tertiary boundary at ODP sites 886 and 803 and DSDP Site 576. Proceeding of the Ocean Drilling Program, Scientific Results 145, 427–434.

Li, L., Keller, G., 1998. Maastrichtian climate, productivity and faunal turnovers in planktic foraminifera in South Atlantic DSDP sites 525 A and 21. Marine Micropaleontology 33, 55–86.

Li, L., Keller, G., 1999. Variability in late cretaceous climate and deep waters: evidence from stable isotopes. Marine Geology 161, 171–190.

Lodders, K., 2003. Solar system abundances and condensation temperatures of the elements. The Astrophysical Journal 591, 1220–1247.

Martic, M., Ajdacic, N., Stjepcevic, J., Gasic, M.J., 1980. Determination of trace elements in marine organisms by neutron activation analysis. Journal of Radioanalytical Chemistry 59, 445–451.

Maruyama, S., Santosh, M., 2008. Models on snowball earth and Cambrian explosion: a synopsis. Gondwana Research 14, 22–32.

Mory, A.J., Iasky, R.P., Glikson, A.Y., Pirajno, F., 2000a. Woodleigh, Carnarvon basin, western Australia: a new 120 km-diameter impact structure. Earth and Planetary Science Letters 177, 119–128.

Mory, A.J., Iasky, R.P., Glikson, A.Y., Pirajno, F., 2000b. Reply to discussion of 'The Woodleigh structure, Carnarvon basin, western Australia: a multi-ring impact structure of 120 km diameter' by U. W. Reimold & C. Koeberl. Earth and Planetary Science Letters 186, 359–365.

Officer, C.B., Drake, C.L., 1985. Terminal cretaceous environmental events. Science 227, 1161–1167.

Pavlov, A.A., Pavlov, A.K., Mills, M.J., Ostryakov, V.M., Vasilyev, G.I., Toon, O.B., 2005a. Catastrophic ozone loss during passage of the solar system through an interstellar cloud. Geophysical Research Letters 32, L01815.

Pavlov, A.A., Toon, O.B., Pavlov, A.K., Bally, J., Pollard, D., 2005b. Passing through a giant molecular cloud: "snowball" glaciations produced by interstellar dust. Geophysical Research Letters 32, L03705.

Planetary Science Institute, 2015. http://www.psi.edu/about/staff/betty/Forcing/.

Poag, C.W., Plescia, J.B., Molzer, P.C., 2002. Ancient impact structures on modern continental shelves: the Chesapeake Bay, Montagnais, and Toms Canyon craters, Atlantic margin of North America. Deep Sea Research Part II: Topical Studies in Oceanography 49, 1081–1102.

Ruderman, M.A., 1974. Possible consequences of nearby supernova explosions for atmospheric ozone and terrestrial life. Science 184, 1079–1081.

Schulte, P., Alegret, L., Arenillas, I., Arz, J.A., Barton, P.J., Bown, P.R., Bralower, T.J., Christeson, G.L., Claeys, P., Cockell, C.S., Collins, G.S., Deutsch, A., Goldin, T.J., Goto, K., Grajales-Nishimura, J.M., Grieve, R.A.F., Gulick, S.P.S., Johnson, K.R., Kiessling, W., Koeberl, C., Kring, D.A., MacLeod, K.G., Matsui, T., Melosh, J., Montanari, A., Morgan, J.V., Neal, C.R., Nichols, D.J., Norris, R.D., Pierazzo, E., Ravizza, G., Rebolledo-Vieyra, M., Reimold, W.U., Robin, E., Salge, T., Speijer, R.P., Sweet, A.R., Urrutia-Fucugauchi, J., Vajda, V., Whalen, M.T., Willumsen, P.S., 2010. The Chicxulub asteroid impact and mass extinction at the Cretaceous–Paleogene boundary. Science 327, 1214–1218.

Shackleton, N.J., Moore Jr., T.C., Rabinowitz, P.D., Boersma, A., Borella, P.E., Chave, A.D., Duee, G., Futterer, D., Jiang, M.-J., Kleinert, K., Lever, A., Manivit, H., O'Connell, S., Richardson, S.H., 1990. Accumulation rates in leg 74 sediments. Initial Reports of the Deep Sea Drilling Project 74, 621–644.

Shukla, P.N., Bhandari, N., Das, A., Shukla, A.D., Ray, J.S., 2001. High iridium concentration of alkaline rocks of Deccan and implications to K/T boundary. Journal of Earth System Science 110, 103–110.

Sloan, R.E., Rigby Jr., J.K., Van Valen, L.M., Gabriel, D., 1986. Gradual dinosaur extinction and simultaneous ungulate radiation in the Hell Creek formation. Science 232, 629–633.

Snoeckx, H., Rea, D.K., Jones, C.E., Ingram, B.L., 1995. Eolian and silica deposition in the central north pacific: results from sites 885/886. Proceeding of the Ocean Drilling Program, Scientific Results 145, 219–230.

Solomon, P.M., Rivolo, A.R., Barrett, J., Yahil, A., 1987. Mass, luminosity, and line width relations of galactic molecular clouds. The Astrophysical Journal 319, 730–741.

Stott, L.D., Kennett, J.P., 1990. The paleoceanographic and paleoclimatic signature of the Cretaceous/Paleogene boundary in the Antarctic: stable isotopic results from ODP leg 113. Proceeding of the Ocean Drilling Program, Scientific Results 113, 829–848.

Sugarman, P.J., Miller, K.G., Burky, D., Feigenson, M.D., 1995. Uppermost Campanian–Maestrichtian strontium isotopic, biostratigraphic, and sequence stratigraphic framework of the New Jersey Coastal Plain. Geological Society of America Bulletin 107, 19–37.

Talbot, R.J., Newman, M., 1977. Encounters between stars and dense interstellar clouds. The Astrophysics Journal Supplement Series 34, 295–308.

Taylor, S.R., McLennan, S.M., 1985. The Continental Crust: Its Composition and Evolution. Blackwell Scientific Publications, Oxford (312 pp.).

Terashima, S., Mita, N., Nakao, S., Ishihara, S., 2002. Platinum and palladium abundances in marine sediments and their geochemical behavior in marine environments. Bulletin of the Geological Survey of Japan 53, 725–747.

Thiede, J., Vallier, T.L., Adelseck, C.G., Boersma, A., Cepek, P., Dean, W.E., Fujii, N., Koporulin, V.I., Rea, D.K., Sancetta, C.A., Sayre, W.O., Seifert, K.E., Schaaf, A., Schmidt, R.R., Windom, K.E., Vincent, E., 1981. Site 463: western mid-Pacific mountains. Initial Reports of the Deep Sea Drilling Project 62, 33–156.

Usui, A., Moritani, T., 1992. Manganese nodule deposits in the central pacific basin: distribution, geochemistry, mineralogy, and genesis. In: Keating, B.H., Bolton, B.R. (Eds.), Geology and Offshore Mineral Resources of the Central Pacific basinCircum-Pacific Council for Energy and Mineral Resources Earth Science Series 14. Springer, New York, pp. 205–223.

Uysal, T., Golding, S.D., Glikson, A.Y., Mory, A.J., Glikson, M., 2001. K–Ar evidence from illitic clays of a late devonian age for the 120 km diameter Woodleigh impact structure, southern Carnarvon basin, western Australia. Earth and Planetary Science Letters 192, 281–289.

Wells, M.C., Boothe, P.N., Presley, B.J., 1988. Iridium in marine organisms. Geochimica et Cosmochimica Acta 52, 1737–1739.

Whitten, R.C., Cuzzi, J., Borucki, W.J., Wolfe, J.H., 1963. Effect of nearby supernova explosions on atmospheric ozone. Nature 263, 398–400.

Wiedmann, J., 1973. Evolution or revolution of ammonoids at mesozoic systemic boundaries. Biological Reviews 48, 159–194.